\documentclass[preprint,10pt,authoryear]{elsarticle}
\usepackage[margin=2.5 cm]{geometry} 

\usepackage{amsmath}
\usepackage{rotating}
\usepackage{multirow}
\usepackage{multicol}
\usepackage{comment}
\usepackage{graphicx}
\usepackage{dcolumn}
\usepackage{natbib}
\usepackage{array,overpic}
\usepackage{amsmath}
\usepackage{booktabs}
\usepackage[colorlinks=true,linkcolor=black, citecolor=cyan, urlcolor=cyan]{hyperref}

\usepackage{bm}
\usepackage{ulem}
\usepackage[dvipsnames]{xcolor}

\def\tensor#1{{\ensuremath{\bm{#1}}}}
\def\vec#1{{\ensuremath{\bm{#1}}}}
\newcommand\p[2]{\ensuremath{\frac{\partial #1}{\partial #2}}}

\def\vec#1{{\ensuremath{{\bm{#1}}}}}
\def\tensor#1{{\ensuremath{{\bm{#1}}}}}

\def\half{{\textstyle \frac{1}{2}}}

\def\p#1#2{{\ensuremath{\frac{\partial #1}{\partial #2}}}}

\newcommand{\hl}[1]{\setlength{\fboxsep}{1pt}\colorbox{white}{#1}}

\renewcommand{\d}{\mathrm{d}}

\newcommand{\jc}[1]{{\emph{\color{orange}{[Jon: #1 ]}}}}

\usepackage{amssymb}

\journal{Journal of the Mechanics and Physics of Solids}

\begin{document}

\begin{frontmatter}



\title{Mechanical stresses in pouch cells: a reduced order model}

\author[inst1]{Andrea Giudici}

\affiliation[inst1]{organization={Mathematical Institute, University of Oxford},
            addressline={Andrew Wiles Building, Woodstock Road}, 
            city={Oxford},
            postcode={OX2 6GG}, 
            state={},
            country={United Kingdom}}

\author[inst1]{Colin Please}
\author[inst1]{Jon Chapman}

\begin{abstract}
In a pouch cell battery, the intercalation of lithium ions into the active particles means the electrodes want to expand. However, since the electrodes are attached to stiff current collectors, this expansion is constrained, leading to a macro-scale deformation and a residually stressed state. This stress state affects the electrochemistry and can also lead  to mechanical degradation, causing a reduction in performance. We model the mechanical state of stress in the battery assuming a known compositional expansion of the electrodes, and use asymptotic techniques to generate reduced order models by exploiting the thin aspect ratio, as well as the large stiffness of the current collectors. We obtain analytic expressions for the stress in the bulk of the electrodes and at the interface between electrodes and current collectors, and a reduced-order equation whose solution  describes the tension in the current collectors. We compare our results with full 3D finite element simulations with excellent agreement, and use our results with the battery simulation package PyBaMM to predict a realistic stress state in a discharging battery.

\end{abstract}

\begin{graphicalabstract}
\includegraphics[width=\textwidth]{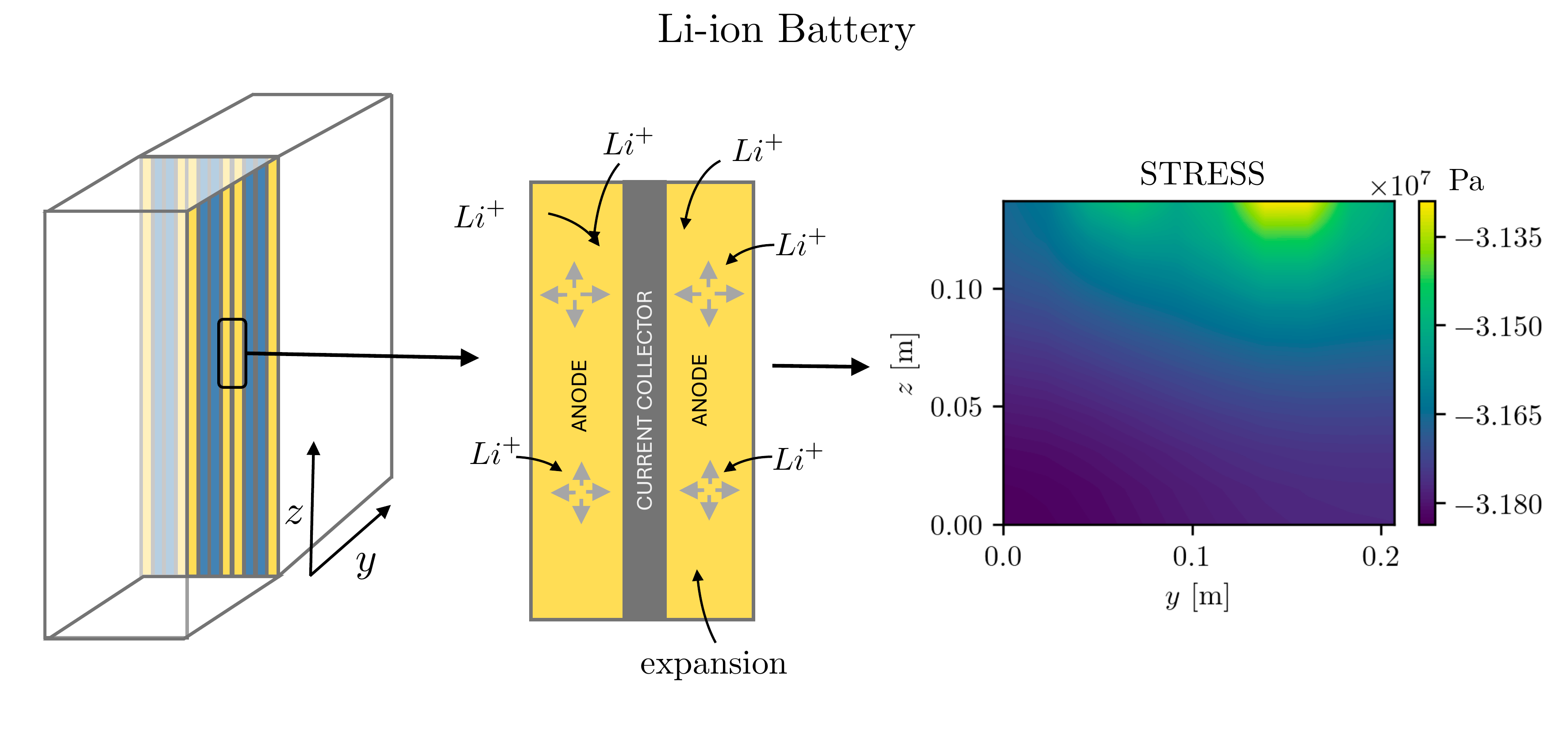}
\end{graphicalabstract}

\begin{highlights}
\item 
The small geometric aspect ratio of the electrodes and vast differences in stiffness between electrodes and current collectors are exploited asymptotically to create a simple model of the mechanical deformation in batteries due to lithiation expansion.

\item 
Comparison between detailed finite element simulations and predictions of the simple model are very good, allowing us to simulate the stress-field in a charging/discharging battery using the battery simulation software PyBaMM.

\end{highlights}


\end{frontmatter}


\section{Introduction}

Mechanical stresses and deformations play an important role in the performance and lifetime of a lithium-ion battery. For example, it is known that operating a battery under light external pressure reduces long term decay and improves capacity \citep{cannarella2014stress}, and cylindrical batteries are known to delaminate and buckle during their operating life causing a reduction in performance \citep{kok2023tracking,waldmann2014mechanical,willenberg2020development,schweidler2018volume}. The origin of mechanical stresses in batteries can primarily be attributed to the movement of lithium ions between electrodes which causes the expansion of the active particles. This expansion results in a variety of mechanically-rich effects at three different scales. 

\begin{figure}[t]
    \centering
    \includegraphics[width=\textwidth]{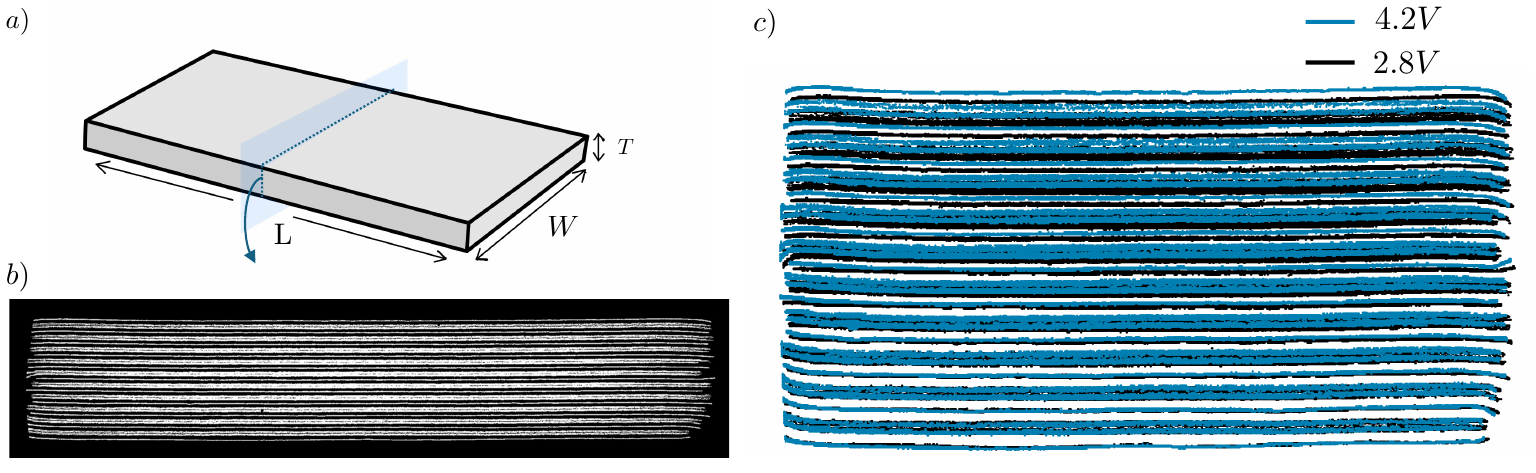}
    \caption{ a) Sketch of a pouch cell. b) X-ray experimental data of the pouch cell showing the different layers. c) Comparison of the X-ray data at $2.8 V$ (fully discharged) and $4.2 V$ (fully charged). The charged state is noticeably expanded compared to the pristine, discharged state, with strains of about $2.5\%$. The aspect ratio has been changed to make the difference more visible. Experimental images were obtained as part of \citep{du2022situ} and are courtesy of Paul Shearing and Wenjia Du. }
    \label{fig:intro}
\end{figure}

At the micro-scale --- the size of the active particles --- expansion and interaction with the surrounding binder material may lead to cracking, delamination and an overall degradation of the active material \citep{ai2022coupled,o2022lithium}. At the meso-scale --- the size of the electrodes --- the particle's expansion means that the active electrode wants to globally expand. The volumetric strain may be as large as $13.2 \%$ in an anode layer \citep{schweidler2018volume}. However, in general, this expansion is limited because the electrode is attached to a stiff current collector (CC) that opposes any deformation, leading to a mismatch in strain. At the macro-scale --- the size of the battery --- this meso-scale mismatch in strain results in deformation of the whole battery, as shown in figure \ref{fig:intro}, with an associated meso-scale state of residual stress. This stress state may be complex, especially when the lithium concentration is not homogeneous across the battery.
Such a scenario is more common when the charge/discharge rates are fast or the conductivity of the electrode is low \citep{timms2021asymptotic}, leading to an inhomogeneous expansion state. 
 
Crucially, the meso-scale stresses depend on the macro-scale deformation and feed back into the micro-scale problem, determining the state of stress that surrounds the active particles. This means that the problems at the three different scales are coupled with each other and require modelling to simulate the mechanical behaviour of a battery. Furthermore, the stress state locally alters the over-potential of the active particle (via an elastic potential term). This change affects the intercalation of lithium, and thus couples the mechanical behaviour with the electrochemistry of the battery.

A model for the meso- and macro-scale mechanics has been proposed for
spirally wound batteries
\citep{timms2023mechanical,psaltis2022homogenization}. Their
non-trivial geometry means that the battery undergoes a complex stress
state during expansion of the active material, with the residual
stresses in the bulk that decrease near the outer and inner
winds. Such a model gives an understanding of the general state of the
system and the regions of where stress concentrates. 

Currently, no meso- and macro- scale mechanical model exists for the
other prominent battery geometry: pouch cells. Pouch cells comprise a
simple layered structure and are commonly used for many applications,
including smart phones and laptops. Experimental observations have
shown that lithiation can cause about a $4\%$ change in thickness of a full cell, suggesting significant stress build-up in the system \citep{leung2014real}. Furthermore, the recent high definition X-ray tomography data of pouch cells in \cite{ du2022situ}
allow us to observe the internal mechanical deformation in great detail, highlighting an overall change in thickness of about $2.5\%$ of the cell form fully discharged to fully charged, as shown in figure \ref{fig:intro}c).


In light of the general interest in understanding the mechanics of batteries, and with the objective of simulating their mechanics efficiently, we propose a reduced-order mathematical model that describes the mechanical state of a pouch cell battery. We exploit the thin aspect ratio of the cell as well as the large stiffness of the current collector to find analytical expressions for the meso-scale stress distribution in the electrodes, the shear stresses at the interface between CCs and electrodes, as well as the tension in the CCs.

The article is structured in the following way: in the first section, we define the geometry of a pouch cell and lay down the assumptions of our model. Most of our assumptions being motivated by observations made in the 3D imagery work presented in  \cite{kok2023tracking} and \cite{du2022situ}.
We then write down the full 3D mechanical equilibrium equations with a prescribed distribution for the expansion of the materials and the boundary conditions. Finally, we non-dimensionalise our equations and note that we cannot, in general, solve the problem analytically. 
In the second section, we proceed to simplify the equilibrium equations by taking the physically appropriate limit in which the electrodes and current collectors are thin compared to their length and width. This approach allows us to reduce our equilibrium equations from full 3D to 2D equations for each electrode CC layer. However, despite the simplification, this problem still cannot be solved analytically. 
In the third section, we therefore take the relevant practical limit in which the stiffness of the CCs is large compared to that of the electrodes. The leading-order solution of this problem is simple and gives us explicit expressions for the stress within the electrode. It also informs us about the shear stresses that are generated at the boundary between CC and electrode due to gradients in the expansion profile. The first order correction yields a problem for the tension in the current collector, which we solve using a Airy stress function and leads to a bi-harmonic equation with a source term which we solve with Dirichlet and Newmann boundary conditions. 
In the fourth section, we show a comparison between our 2D reduced order model and full 3D Finite Element simulation and comment on the good agreement between the two. Furthermore, using PyBaMM \citep{sulzer2021python}, an open source battery modelling library in Python, we simulate the electrochemical behaviour of a battery during fast discharging and use it to derive the expansion that occurs in the electrode materials. We then use our model to determine the stress field and mechanical state of the system.
Finally, in the fifth section, we make our concluding remarks.

\section{Geometry of pouch cell and problem set up}

\begin{figure}[h!]
    \centering
    \includegraphics[width=\textwidth]{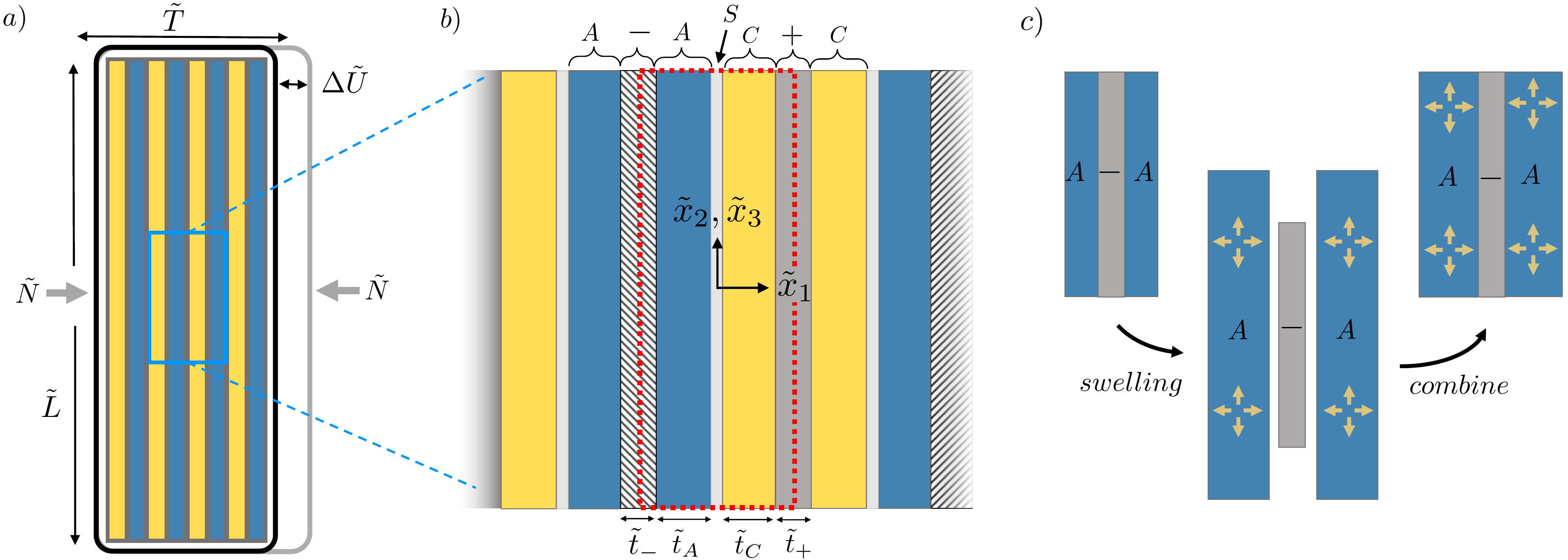}
    \caption{ a) Schematic of a pouch cell of thickness $\tilde{T}$, length $\tilde{L}$ and width $\tilde{W}$ (not shown, into the page), with a normal force $\tilde{N}$ constraining the system and an associated relaxation displacement $\Delta\Tilde{ U}$. b) A closer view of the layered structure within the battery. The unit cell, up to mirror symmetry, is shown in dashed red. c) Schematic of the mismatch in strain generated by lithiation-induced swelling in the anode. If detached, the anode and the current collector would relax to different lengths (second panel). However, since they are attached and the CC is stiff, the length of electrode in contact with the CC is constrained, resulting in the build up of internal stress.  }
    \label{fig:PouchSketch}
\end{figure}

\subsection{Introduction of the pouch cell model}

We consider a pouch cell of thickness $\tilde{T}$, length $\tilde{L}$
and width $\tilde{W}$, composed of a layered structure of alternating
cathode (C) and anode (A) layers with their respective current
collectors ($+,-$), separated by a thin and soft separator (S). We
define an anode pair as the layer composed of the two anode electrodes and the $(-)$
CC collector in the middle. Similarly, the cathode
pair is composed of 
the two cathode layers and the $(+)$ CC in the middle. There are $n$
cathode and anode pairs which we assume to be identical to each other
up to mirror symmetry. A schematic of the pouch cell is shown in
figure \ref{fig:PouchSketch} a) and b). Due to swelling, the pouch
cell may expand along the through-cell direction $\tilde{x}_1$. The
expansion may be constrained by hard clamps, fixing the magnitude of
$\Delta \tilde{ U}$, or may be opposed by an external load
$\tilde{N}$. The two cases are mathematically related and we shall
consider both.

During a charge cycle, the cathode and anode undergo different isotropic expansion with a (linearised) volumetric strain $3 \alpha_C(x_1,x_2,x_3)$ and $3 \alpha_A(x_1,x_2,x_3)$ respectively. In contrast, the current collectors do not undergo such expansion.
This process of differential swelling generates a mismatch in strain and frustration at the interface between electrode and current collector. If the system was a sheet composed of only two layers, say anode and CC, the sheet would bend with curvature proportional to the mismatch in strain, just as a bimetallic strip would \citep{timoshenko1925analysis}. This bending effect has, in fact, been used to measure the mechanical properties of swelling electrodes, such as their electrochemical stiffness \citep{tavassol2016electrochemical}. However, since the electrodes are on both sides of the CC, no bending is induced and the mismatch in strain is partially relaxed via Poisson effects only, as shown schematically in figure \ref{fig:PouchSketch} c). 

Locally, we may still expect bending to arise as a result of inhomogeneities in the swelling of the electrodes. For instance, if the electrodes were to swell much more in the middle of the structure than they do at the boundary, one may expect the CC to bend and the whole structure to bulge out. Although this bulging can be observed in batteries, it usually does not occur as a result of lithiation-induced expansion, but rather as a consequence of build-up of gas within the cell causing large pressures and degradation of the material \citep{du2022situ}. 

In the case of lithium induced expansion, the strains are relatively small and since the CC are stiff, any bulging would also cause a change in the Gaussian curvature of the CC sheet, causing in-plane stretch with prohibitive energetic costs. Furthermore, batteries tend to be operated within a casing that clamps the outer surfaces, forcing them to be flat.  Therefore, we shall assume that the mid-planes of each current collector remain completely flat and parallel to each other, while still allowing the CCs to move with respect to each other to find mechanical equilibrium. In light of our assumption, we shall consider the unit cell shown in dashed red in figure \ref{fig:PouchSketch} b). Our objective is to understand how the cell copes with the differential expansion in the active material, how it deforms, and describe the resulting residual stresses  in the active material, in the current collectors, and at their interfaces.

Finally, it is important to comment on the coupling between electrochemical and mechanical effects. Indeed, in general, the mechanical stresses affect the local open circuit potential of the electrode and thus the local lithium concentration. Since the lithium concentration is responsible for the expansion (and thus the stress) in the first place, there is in fact a two-way coupling.  Here, we only consider and solve the one way coupling problem in which, for a given lithium concentration, we find the mechanical deformation and stress state. Since the time-scale of elastic relaxation is much shorter than the timescale of  battery charging/discharging (even at high C-rates), our analysis essentially provides the quasi-steady state relationship between lithium concentration and stress, which can then be used as part of a fully coupled dynamic electro-chemical-mechanical simulation.

\subsection{Full 3D problem: governing equations}
\label{sec:3Dproblem_setup}

We model the layers as linear elastic solids undergoing isotropic expansion so that, if a small piece of material was allowed to deform unconstrained, its strain would be $\tensor{\tilde{\epsilon}}=\tilde{\alpha}_k \tensor{I}$, where $\tensor{I}$ is the identity matrix, $k \in \{+,-,A,C\}$ identifies the different layers (positive current collector, negative current collector, anode and cathode respectively) and $\tilde{\alpha}_k$ is one-third of the volumetric strain, i.e.~it is the linear expansion coefficient of the material. We shall model the separator differently, and introduce this later in the following derivation. Note that we are labelling here all quantities with an overtilde which we shall drop when we later non-dimensionalise the system. Furthermore, note that $\tilde{\alpha}_k=0$ when $k \neq A,C$ since the current collectors are assumed to remain passive.

We describe the deformation in the materials via the general displacement field
\begin{equation}
    \vec{\tilde{u}}^k=(\tilde{u}^k_1(\tilde{x}_1,\tilde{x}_2,\tilde{x}_3),\tilde{u}^k_2(\tilde{x}_1,\tilde{x}_2,\tilde{x}_3),\tilde{u}^k_3(\tilde{x}_1,\tilde{x}_2,\tilde{x}_3)).
\end{equation}
where once again $k$ indicates which layer the displacement refers to. This displacement field gives rise to a strain (with respect to the reference configuration) of the form:
\begin{equation}
\label{eq:strain}
    \tilde{\epsilon}^k_{ij}=\frac{1}{2} \left(\p {\tilde{u}^k_j}{\tilde{x}_i}+\p {\tilde{u}^k_i}{\tilde{x}_j} \right).
\end{equation}
 Since we are dealing with linear elasticity, we may identify the effective strain in the material as the strain induced by the deformation minus the strain preferred by the expansion:
\begin{equation}
    \tilde{\varepsilon}^k_{ij}=\tilde{\epsilon}^k_{ij}-\tilde{\alpha}^k \delta_{ij}.
\end{equation}
where $\delta_{ij}$ is the Kronecker delta.
For a material of Young's modulus $\tilde{E}_k$ and Poisson ratio $\nu_k$, the resulting stress in the material is
\begin{align}
\label{eqn:stress}
\tilde{\sigma}^k_{ij}=\frac{\tilde{E}_k}{(1 + \nu_k)}\left[\tilde{\varepsilon}^k_{ij}+\frac{\nu_k}{(1-2 \nu_k)}\tilde{\varepsilon}^k_{ll} \delta_{ij}\right].
\end{align}
It is also convenient to introduce the in-plane tension in a layer. This can be obtained by integrating the in-plane component of stress over the whole thickness of the layer (which we identify as $\int_{\tilde{t}_k}d\tilde{x}_1$). Letting, from now on, Greek subscripts $\eta, \zeta =\{2,3\}$ identify only in-plane components, the tension is:
\begin{align}
\label{def:tension}
    \tilde{T}^k_{\eta \zeta}=\int_{\tilde{t}_k} \tilde{\sigma}^k_{\eta
      \zeta}\; \d \tilde{x}_1.
\end{align}
Finally, we generalise this concept by defining the average over the thickness of a general quantity $\psi_k$ in  layer $k$ as
\begin{equation}
    \langle \psi_k \rangle = \frac{1}{\tilde{t}_k}\int_{\tilde{t}_k}
    \psi_k \,\d \tilde{x}_1 \label{thicknessaverage}
\end{equation}
and the global average, over the entire cell in layer $k$, as
\begin{equation}
    \overline{\langle \psi_k \rangle} =
    \frac{1}{\tilde{W}\tilde{L}}\int_0^{\tilde{L}} \int_0^{\tilde{W}}
    \langle \psi_k \rangle \,\d \tilde{x}_2 \,\d \tilde{x}_3.
\end{equation}

\subsection{Elastic equilibrium equations and boundary conditions}
The equilibrium conditions require the divergence of the stress tensor to vanish, so that
\begin{align}
    \p{}{\tilde{x}_j} \tilde{\sigma}^k_{ij}=0,
    \label{eqn:equilibrium}
  \end{align}
where here, and henceforth, we use Einstein's summation convention. These equations are accompanied by matching conditions between layers and appropriate boundary conditions  on the unit cell boundaries. Since the electrodes are attached to their respective current collector, we require continuity of displacement and stress across their interface, meaning
\begin{align}
\label{bc:continuity1}
\left[ \vec{\tilde{u}} \right]_-^+&=\left[ \tilde{u}_i \right]_-^+=0\quad\,\,\,\text{at} \,\,\,\tilde{x}=\tilde{t}_C,\\
\label{bc:continuity2}
\left[ \vec{\tilde{\sigma}} \cdot \vec{\hat{x}}_1 \right]_-^+&=\left[ \tilde{\sigma}_{1j}  \right]_-^+=0\quad\,\,\,\text{at} \,\,\,\tilde{x}=\tilde{t}_C,\\
\label{bc:continuity3}
\left[ \vec{\tilde{u}} \right]_-^+&=\left[ \tilde{u}_i \right]_-^+=0\quad\,\,\,\text{at} \,\,\, \tilde{x}=-\tilde{t}_A,\\
\label{bc:continuity4}
\left[ \vec{\tilde{\sigma}} \cdot \vec{\hat{x}}_1 \right]_-^+&=\left[ \tilde{\sigma}_{1j}  \right]_-^+=0\quad\,\,\,\text{at} \,\,\, \tilde{x}=-\tilde{t}_A,
\end{align}
where $[\,\,]_-^+$ is the jump across the interface, and
$\vec{\hat{x}}_1$ is the unit vector in the $\tilde{x}_1$ direction.

We assume that the separator is much thinner and softer than the electrodes. Furthermore, we assume that the battery is under compression, so that the displacements in the separator may be at most as large as its thickness and therefore vanishingly small compared to the electrode thickness. As a result, when considering the through-cell direction, the electrodes effectively behave as if there was no separator in between, allowing us to impose continuity of through-cell displacement and stress.  Conversely, since the separator is extremely soft, it offers no resistance to shear strains, and we therefore assume no shear stresses. Effectively, the boundary between electrodes behaves as a sliding contact. Letting Greek subscripts $\eta,\zeta$ represent only elements in the $\tilde{x}_2,\tilde{x}_3$ plane so that $\eta,\zeta \in\{2,3\}$, we can write the boundary conditions on the separator as:
\begin{align}
\label{bc:separator1}
\tilde{u}^C_1 &=\tilde{u}^A_1\quad\,\,\,\text{at} \,\,\,\tilde{x}_1=0,\\
\label{bc:separator2}
\tilde{\sigma}^C_{11} &= \tilde{\sigma}^A_{11} \quad\,\,\,\text{at} \,\,\,\tilde{x}_1=0,\\
\label{bc:separator3}
\tilde{\sigma}^A_{1\eta }&=0\quad\,\,\,\text{at} \,\,\, \tilde{x}_1=0,\\
\label{bc:separator4}
\tilde{\sigma}^C_{1\eta }&=0\quad\,\,\,\text{at} \,\,\, \tilde{x}_1=0,
\end{align}
where the latter two represent the vanishing of normal shear stresses at the interface.

The boundary conditions along the mid-plane of the CCs arise from left-right symmetry, leading to:
\begin{align}
\label{bc:symmetry1}
   \p{\tilde{u}_{\eta}^+}{\tilde{x}_1}&=0\quad\,\,\,\text{at} \,\,\, \tilde{x}_1=-(\tilde{t}_A+ \half \tilde{t}_-),\\
\label{bc:symmetry2}
     \p{\tilde{u}_{\eta}^-}{\tilde{x}_1} &=0\quad\,\,\,\text{at} \,\,\, \tilde{x}_1=\tilde{t}_C+\half  \tilde{t}_+.
\end{align}
Furthermore, as discussed, we require the mid-planes of the current collectors to remain flat. Therefore, without loss of generality, we choose
\begin{equation}
\label{bc:u}
\tilde{u}_1(-\tilde{t}_A-\half \tilde{t}_-,x_2,x_3)=0, \quad \tilde{u}_1(t_C+\half \tilde{t}_+,x_2,x_3)=\Delta \tilde{u} 
\end{equation}
where $\Delta \tilde{ u}=\Delta \tilde{ U}/n$ is the constant displacement between the two CCs, which is equal to the total change in thickness in the battery divided by the number of unit cells (or, equivalently, divided by the number of  electrode and CC pairs). This displacement may be a constraint, coming from fixed clamps, or be the result of a force balance found by a total load $\tilde{N}$ applied on the ends of the system. We shall concentrate on the case where force balance is relevant and then the global force balance requires that 
\begin{align}
\label{bc:global}
  \int_{-\tilde{W}/2}^{\tilde{W}/2} \int_{-\tilde{L}/2}^{\tilde{L}/2} (\vec{\hat{x}_1} \cdot \tensor{\tilde{\sigma}} \cdot \vec{\hat{x}}_1)\; d\tilde{x}_2 d\tilde{x}_3 &=\tilde{N}\quad\,\,\,\text{at} \,\,\, \tilde{x}_1=\pm \tilde{T}/2,
\end{align}
where \(\tilde{T}\) is the macro-scale thickness of the battery. This equation connects the meso-scale mechanics to the macro-scale loads applied to the battery.

Note that a compressive load has negative sign, $\tilde{N}<0$. Since load is usually imposed by clamping the battery with springs or by applying an external pressure $\tilde{p}$ (so that \(\tilde{N}=-\tilde{p} \tilde{L} \tilde{W}\) where negative sign ensures the pressure is positive in compression).
Finally, in the absence of tangential forces, we also require the stress at the sides of the layers to vanish, meaning
\begin{align}
\label{bc:side1}
    \tensor{\tilde{\sigma}}\cdot \hat{\vec{x}}_2&=\tilde{\sigma}_{2j}=0\quad\,\,\,\text{at} \,\,\, \tilde{x}_2=\pm \tilde{L}/2,\\
\label{bc:side2}
    \tensor{\tilde{\sigma}}\cdot \hat{\vec{x}}_3&=\tilde{\sigma}_{3j}=0\quad\,\,\,\text{at} \,\,\, \tilde{x}_3=\pm \tilde{W}/2.
\end{align}
where $j \in \{1,2,3 \}$.

\subsection{Non-dimensionalisation}

We non-dimensionalise the problem and rescale each quantity by its typical size. We let $L^*$ be the typical length (and width) of the battery and $t^*$ the typical thickness of the layers. Letting the typical size of swelling be $\alpha^*$, our non-dimensional swelling and lengths can be expressed as:
\[
    \tilde{\alpha}_k=  \alpha^* \alpha_k(x_1,x_2,x_3),\quad
    \tilde{x}_1= t^* x_1,\quad 
    \tilde{x}_2= L^* x_2,\quad
    \tilde{x}_3= L^* x_3,\quad
    \tilde{u}^k_{1}= \alpha^*  t^*  u^k_{1},\quad 
    \tilde{u}^k_{\eta}=\alpha^* L^* u^k_{\eta},\]
  \[
   \Delta \tilde{u}_1=t^*  \Delta u , \quad  \tilde{t_k}= t^* t_k,\quad
    \tilde{L}= L^* L,\quad
    \tilde{W}= L^* W, \quad \tilde{T}=t^* T,
\]
where once again the index $k$ identifies the layer and $\eta \in \{2,3\}$ selects only in-plane components.

We also rescale the strains, stresses and loads in the system the typical Young's modulus of the electrodes $E^*$, so that 
\begin{equation}
    \tilde{\epsilon}^k_{ij}=  \alpha^* \epsilon^k_{ij},\quad\quad\tilde{\sigma}^k_{ij}= \alpha^* E^* \sigma^k_{ij},\quad\quad\tilde{N}=  \alpha^* E^* {L^*}^2 N,\quad\quad\tilde{T}_{\eta\zeta}= \alpha^* E^* t^* T_{\eta\zeta},
\end{equation}
where we have that $\tilde{E}_k=E^* E_k$.

Our elastic problem can thus be re-written entirely in dimensionless form by simply dropping all tildes from the equilibrium equations and the boundary conditions in the previous section. The only difference is that all the $u_1$ fields now pick a prefactor of $\delta$ while derivatives with respect to $\tilde{x}_1$ in equations \eqref{eq:strain},\eqref{eqn:equilibrium},\eqref{bc:symmetry1} and \eqref{bc:symmetry2}, pick a prefactor of $\delta^{-1}$. We shall spell these equations out more clearly in the next section, where we expand our fields in small thickness $\delta \ll 1$. 
For any lithiation expansion function $\alpha_k$, we can now in
principal determine the stress state and deformation in the battery. However, in general, this problem is complex and can only be solved numerically, which incurs significant computational cost. To make analytical progress, reduce computational cost, and gain understanding, we shall consider the physically relevant limit where the electrodes and CCs are thin and where the CCs are much stiffer than the electrodes.

\section{Thin aspect ratio asymptotics}

We look at the limit in which the layers are thin compared to their
length and width, that is, we assume that the typical thickness $t^*$
of a layer is much smaller than its  typical length (or width) $L^*$, meaning that the parameter $\delta=t^*/L^* \ll 1$.
Before we proceed to expand the fields and solve our problem in this limit, we remark that one significant difficulty that arises in this analysis comes from the boundary conditions \eqref{bc:side1} and \eqref{bc:side2}. Near this outer boundary, there is a local deformation --- a boundary layer --- that needs to be solved for separately. This is beyond the scope of this work. However, we can still make progress by studying the solution away from this boundary in the bulk of the cell (the ``outer region'' of the asymptotic expansion). As we shall see, for this outer region the edge boundary conditions \eqref {bc:side1} and \eqref{bc:side2} will
translate to a condition for the CC electrode pair integrated over the thickness.  

We now proceed with our asymptotic expansion. We write the displacement fields as series expansions in small $\delta$,
\begin{equation}
    u^k_{i}=    u^{k(0)}_{i} + u^{k(1)}_{i} \delta +u^{k(2)}_{i} \delta^2+ \dots,
\end{equation}
this leads to an expansion for the strain tensor in equation \eqref{eq:strain} given by
\begin{align}
\varepsilon^k_{11}&=   \varepsilon^{k(0)}_{11}+\varepsilon^{k(1)}_{11} \delta + \varepsilon^{k(2)}_{11} \delta^2 +\dots,\\
\varepsilon^k_{1 \zeta}&=   \varepsilon^{k(-1)}_{1 \zeta} \delta^{-1}+\varepsilon^{k(0)}_{1 \zeta} + \varepsilon^{k(1)}_{1 \zeta} \delta +\dots,\\
\varepsilon^k_{\eta \zeta}&=   \varepsilon^{k(0)}_{\eta \zeta}+\varepsilon^{k(1)}_{\eta \zeta} \delta + \varepsilon^{k(2)}_{\eta \zeta} \delta^2 \dots, \quad \eta, \zeta \in \{2,3\}.
\end{align}
The stress in equation \eqref{eqn:stress} has an expansion of the form
\begin{align}
\sigma^k_{11}&=   \sigma^{k(0)}_{11}+\sigma^{k(1)}_{11} \delta + \sigma^{k(2)}_{11} \delta^2 +\dots,\\
\sigma^k_{1 \zeta}&=   \sigma^{k(-1)}_{1 \zeta} \delta^{-1}+\sigma^{k(0)}_{1 \zeta} + \sigma^{k(1)}_{1 \zeta} \delta +\dots,\\
\sigma^k_{\eta \zeta}&=   \sigma^{k(0)}_{\eta \zeta}+\sigma^{k(1)}_{\eta \zeta} \delta + \sigma^{k(2)}_{\eta \zeta} \delta^2 \dots, \quad \eta, \zeta \in \{2,3\}.
\end{align}
Note that we have allowed the strains and stresses with components along the the $x_1$ direction to be of order $\delta^{-1}$. These terms arise due to derivatives in the $x_1$ direction of displacements $u_2$ and $u_3$. We can now expand the equilibrium conditions \eqref{eqn:equilibrium} and derive new and simpler equations for the leading-order stress. Later, we shall expand the stress as a function of displacements and write down the full problem for the displacements.

Since derivatives with respect to $x_1$ yield a $\delta^{-1}$ factor, at leading order we have the equation
\begin{equation}
\label{eq:-1B}
    \partial_1 \sigma^{k(-1)}_{1j}=0
    \qquad j \in \{1,2,3\},
  \end{equation}
  where $\partial_i \equiv \partial/\partial x_i$.
There is no term $\sigma^{k(-1)}_{11}$, meaning equation \eqref{eq:-1B} is already satisfied when $j=1$.
  When $j=\zeta \in \{2,3\}$, we have that $\sigma^{k(-1)}_{1 \zeta}$ is constant. Applying the no shear boundary conditions at the separator \eqref{bc:separator3} and \eqref{bc:separator4}, we deduce that $\sigma^{k(-1)}_{1 \zeta}=0$, meaning all terms of order $\delta^{-1}$ in the stress vanish as we would expect. 

At the next order we have that
\begin{equation}
    \partial_1 \sigma^{k(0)}_{1j}+\partial_{\zeta} \sigma^{k(-1)}_{j\zeta}=0,
    \qquad j \in \{1,2,3\},\;\zeta \in \{2,3\}.
    \label{eq:equilibriumexpandedstress0}
\end{equation}
However, since all components of the stress tensor $\tensor{\sigma}^{k(-1)}$ vanish, we obtain
\begin{equation}
\label{eq:-1}
    \partial_1 \sigma^{k(0)}_{11}=0, \quad \quad \partial_{1}
    \sigma^{k(0)}_{1\zeta}=0. 
\end{equation}
The first equation tells us that the stress is (piece-wise) constant in $x_1$. Continuity of normal stress across all interfaces,
\eqref{bc:continuity2} and \eqref{bc:continuity4},  requires a
constant normal stress through the layer, which allows us to write
\begin{equation}
\label{eq:cond_sigma11}
 \sigma^{k(0)}_{11}=c(x_2,x_3).
\end{equation}
At the next order, the equilibrium condition reads
\begin{equation}
    \partial_1 \sigma^{k(1)}_{1j}+\partial_{\zeta} \sigma^{k(0)}_{j\zeta}=0
    \qquad j \in \{1,2,3\},\;\zeta \in \{2,3\}.
    \label{eq:equilibriumexpandedstress}
\end{equation}
This equation relates the leading-order  $x_2$ and $x_3$ components of stresses to the first order $x_1$ shears. The first order shears can
be eliminated from the above equation by making use of the boundary
conditions. We therefore integrate
\eqref{eq:equilibriumexpandedstress} over the thickness to give
\begin{equation}
    \partial_{\zeta} T^{k(0)}_{ \eta \zeta}=-\left[\sigma^{k(1)}_{1 \zeta}\right]_{t_k}
    \qquad \eta,\zeta \in \{2,3\}.
    \label{eq:equilibriumO1}
\end{equation}
where $T^{k(0)}_{ \eta \zeta}$ is the component of the tension defined in \eqref{def:tension} and
$[\cdot]_{t_k}$ is the jump across layer $k$ (of thickness $t_k$), so
that, for instance, $\left[\sigma^{A(1)}_{1 \zeta}\right]_{t_A}=\sigma^{A(1)}_{1 \zeta}|_{x_1=0}-\sigma^{A(1)}_{1 \zeta}|_{x_1=-t_A}$. This equation suggests that the leading-order in-plane tension is generated by the difference in first order shear stresses at the two interfaces of the electrode.

\subsection{Global equilibrium equation}
We know that stresses are continuous across the interface between each
CC and its adjacent electrode. We can exploit this property to
simplify  \eqref{eq:equilibriumO1}. The approach involves
introducing the total tension in a electrode and current collector pair, composed of the current collector and the two electrodes on either
side of it. However, the unit cell we are considering contains only
half of each electrode pair, i.e. half of each CC together with the
corresponding electrode. We therefore define the tensions in each half-electrode CC pair, that is 
\begin{align}
\label{def:Sigma}
   \Sigma^+_{\eta \zeta}&\equiv \half T_{\eta \zeta}^+ +
   T_{\eta \zeta}^C ,\qquad \eta,\zeta \in \{2,3\},\\
   \Sigma^-_{\eta \zeta}&\equiv \half T_{\eta \zeta}^- +
   T_{\eta \zeta}^A ,\qquad \eta,\zeta \in \{2,3\}.
\end{align}

Let us now consider only the cathode pair, with the calculation being equivalent for the anode pair. We assume the expansion
$\Sigma^+_{\eta \zeta}= \Sigma^{+(0)}_{\eta \zeta} +   \Sigma^{+(1)}_{\eta \zeta} \delta +\Sigma^{+(2)}_{\eta \zeta} \delta^2+...\,$.
Using equation \eqref{def:tension} and taking the divergence of $\Sigma^+_{\eta \zeta}$ in the $x_2$, $x_3$ plane, at leading order we find that
\begin{align}
\partial_{\zeta}\Sigma^{+(0)}_{\eta \zeta}&=\int_{0}^{t_C}
\partial_{\zeta}\sigma^{C(0)}_{\eta \zeta}\; \d x_1+\int^{t_C+\half
  t_+}_{t_C} \partial_{\zeta}\sigma^{+(0)}_{\eta \zeta}\; \d x_1\\
\label{eq:eq2}
&=-\int_{0}^{t_C} \partial_{1}\sigma^{C(1)}_{1 \zeta}\; \d
x_1-\int^{t_C+\half t_+}_{ t_C} \partial_{1}\sigma^{+(1)}_{1 \zeta}\;
\d x_1
\end{align}
where the second equality was obtained using equation
\eqref{eq:equilibriumexpandedstress}. We can now evaluate the
integrals  in $x_1$ and use continuity of stress across all interfaces, equation \eqref{bc:continuity2}, to obtain
\begin{align}
\partial_{\zeta} \Sigma^{+(0)}_{{\eta} \zeta}=-\left.\sigma^{C(1)}_{1 \zeta}\right|_{x_1=0}-\left.\sigma^{+(1)}_{1 \zeta}\right|_{x_1={t_C+\half t_+}}.
\end{align}

Finally, since the separator does not transmit shear stresses,
equations \eqref{bc:separator3} and \eqref{bc:separator4} imply that
$\sigma^{C(1)}_{1 \zeta}|_{x_1=0}=\sigma^{A(1)}_{1 \zeta}|_{x_1=0}=0$. Similarly,
by symmetry across the mid-plane of the current collector, we have
$\sigma^{-(1)}_{1 \zeta}|_{x_1=t_C+\half
  t_+}=\sigma^{+(1)}_{1 \zeta}|_{x_1=-t_A-\half t_-}=0$, where $\eta \in \{2,3\}$.
  Putting everything together we can write the equilibrium equations as
\begin{align}
\label{eq:equilibriumF}
\partial_{\zeta} \Sigma^{+(0)}_{\eta \zeta}&=\partial_{\zeta} \left(T^{C(0)}_{\eta \zeta}+\half T^{+(0)}_{\eta \zeta} \right)=0,\\
\partial_{\zeta} \Sigma^{-(0)}_{\eta {\zeta}}&=\partial_{\zeta} \left(T^{A(0)}_{\eta \zeta}+\half T^{-(0)}_{\eta \zeta} \right)=0\,,\quad\,\,\,\eta,\, \zeta \in \{2,3\}.
\label{eq:equilibriumF2}
\end{align}

Defining the total tension $\tensor{\Sigma}$ has allowed us to write
our 3D problem as two coupled 2D problems for each electrode and CC
pair. Furthermore, since the total tension $\tensor{\Sigma}$ is an
integral over the thickness, we can now express the boundary
conditions, \eqref{bc:side1} and \eqref{bc:side2}, as global integrals
over the thickness too, 
\begin{align}
\label{bc:side1B}
    \tensor{\Sigma}^{-(0)} \cdot \vec{\hat{x}}_2&=0, \quad
    \tensor{\Sigma}^{+(0)} \cdot \vec{\hat{x}}_2=0,\qquad \mbox{ on }
    x_2 = \pm L/2,\\
    \label{bc:side2B}
    \tensor{\Sigma}^{-(0)} \cdot \vec{\hat{x}}_3&=0, \quad
    \tensor{\Sigma}^{+(0)} \cdot \vec{\hat{x}}_3=0,\qquad \mbox{ on }
    x_3 = \pm W/2,
\end{align}
thus avoiding  having to solve the boundary layer problem. We can now use our simplified equilibrium condition to write down the whole problem for the leading-order displacements. 

\subsection{Full problem for the leading-order displacement}

The leading-order displacements in the system are $ u_1^{k(0)}, u_2^{k(0)}, u_3^{k(0)}$. 
We use the constitutive relations \eqref{eqn:stress} to write down the relevant entries of the stress tensor:
\begin{align}
\label{stressExpansion1}
     \sigma^{k(-1)}_{1 \zeta}&=\frac{E_k}{(1+\nu_k)}  \p{u_{\eta}^{k(0)}}{x_1}\\
     \label{stressExpansion2}
     \sigma^{k(0)}_{11}&=\frac{E_k}{(1+\nu_k)} \left[ \frac{(1-\nu_k)}{(1-2 \nu_k)} \p{u_1^{k(0)}}{x_1}+\frac{\nu_k}{(1-2 \nu_k)} \p{u_{\beta}^{k(0)}}{x_{\beta}} \right] - \frac{E_k}{(1-2 \nu_k)} \alpha_k \\
     \label{stressExpansion3}
     \sigma^{k(0)}_{\eta \zeta }&=\frac{E_k}{(1+\nu_k)} \left[
       \frac{1}{2}\left(\p{u_{\zeta}^{k(0)}}{x_{\eta}}+\p{u_{\eta}^{k(0)}}{x_{\zeta}}\right)+\frac{\nu_k}{(1-2
         \nu_k)} \left(
         \p{u_{\beta}^{k(0)}}{x_{\beta}}+\p{u_1^{k(0)}}{x_1}\right)
       \delta_{\eta \zeta} \right]\nonumber  \\
     & \hspace{8cm}- \frac{E_k}{(1-2 \nu_k)} \alpha_k \delta_{\eta \zeta}, \quad \eta, \zeta \in \{2,3\},
\end{align}
where for the term ${\partial u_{\beta}^{k(0)}}/{\partial x_{\beta}}$ summation is over indices $\beta \in \{2,3\}$. From \eqref{eq:-1} we know that $\sigma^{k(-1)}_{1 \zeta}=0$.
 Therefore, we deduce that the in-plane displacements are independent of the through-cell direction, meaning
\begin{equation}
u^{k(0)}_{\eta}=u^{k(0)}_{\eta}(x_2,x_3).
\end{equation}
Furthermore, the continuity of displacement across the interface between current collector and electrode, \eqref{bc:continuity1} and \eqref{bc:continuity3}, tell us that 
\begin{equation}
\label{condition_u}
u^{C(0)}_{\eta}=u^{+(0)}_{\eta} , \quad \quad u^{A(0)}_{\eta}=u^{-(0)}_{\eta}.
\end{equation}
These simplifications have a few important benefits. First, because
the in-plane fields are independent of the through-cell direction
$x_1$, we can easily integrate them over the thickness --- they simply
get multiplied by $ t_k$. Similarly, the integral of the through-cell
displacement will give us the total  displacement (change in thickness)
of the layer.

We now drop all the superscript $(0)$ since it is clear we are
evaluating the leading-order term for every quantity. Then,
integrating the stresses in \eqref{stressExpansion3} over the
thickness, we find the leading-order term in the tension is
\begin{equation}
\label{eq:tension_leading_order}
    T^{k}_{\eta \zeta }=\frac{ E_k t_k}{(1+\nu_k)} \left[ \frac{1}{2}\left(\p{u_{\zeta}^{k}}{x_{\eta}}+\p{u_{\eta}^{k}}{x_{\zeta}}\right)+\frac{\nu_k}{(1-2 \nu_k)} \left( \p{u_{\beta}^{k}}{x_{\beta}}+\frac{\Delta u^{k}_1}{t_k} \right) \delta_{\eta \zeta} \right] - \frac{E_k }{(1-2 \nu_k)} \alpha_k \delta_{\eta \zeta}, \quad \eta, \zeta \in \{2,3\},
\end{equation}
  where $\Delta u^{k}_1=\int_{t_k} \p{u^{k}_1}{x_1} dx_1$ is the leading-order total through-cell displacement (i.e. the change in thickness) of the layer. We can substitute these definitions of the tension in \eqref{def:Sigma} to find the total tension in the electrode CC pair. 

Overall, our problem has 9 unknown functions to solve for, the displacements $u^{C}_1$, $u^{C}_2$, $u^{C}_3$, $u^{+}_1$,$u^{A}_1$, $u^{A}_2$, $u^{A}_3$, $u^{-}_1$ and the through-cell stress $c(x_2,x_3)$.
Four equations for the tension come from the equilibrium conditions, \eqref{eq:equilibriumF} and \eqref{eq:equilibriumF2} with boundary conditions \eqref{bc:side1B} and \eqref{bc:side2B}. These equations relate the divergences of the tension in a pair of electrode and current collector.  
The relationship between the through-cell stress and the through-cell displacements comes from \eqref{eq:cond_sigma11} and yields 4 more differential equations for the $u_1^k$'s,
\begin{equation}
\label{equation_u1}
    \frac{E_k}{(1+\nu_k)} \left[ \frac{(1-\nu_k)}{(1-2 \nu_k)} \p{u_1^{k}}{x_1}+\frac{\nu_k}{(1-2 \nu_k)} \left( \p{u_2^{k}}{x_2}+\p{u_3^{k}}{x_3}\right) \right] - \frac{E_k}{(1-2 \nu_k)} \alpha_k=c(x_2,x_3)
\end{equation}
where the boundary conditions are just continuity of $u_1$ across all interfaces, equations \eqref{bc:continuity1}, \eqref{bc:continuity3}, \eqref{bc:separator1}, and condition \eqref{bc:u}. The problem is closed either by fixing the total displacement $\Delta u_1 = \Delta U/n$, or using equation \eqref{bc:global} to find the relationship between the total displacement $\Delta U$ and the applied load $N$.

In the thin limit, the mechanical state of the system is therefore described by two 2D problems for the in-plane displacements coupled via the through-cell stress (coming form the term $c(x_2,x_3)$).
For given swelling parameters and material properties, this problem can be solved numerically with reduced computational cost compared to the original full 3D one.
However, in general, it still cannot be solved analytically. To make analytical progress and simplify the computation further, we look at the limit in which the current collectors are stiff compared to the electrode.

\section{Stiff Current Collector limit}
\label{sec:stifflimit}

Here, we solve the leading-order problem derived in the previous section for the case in which the current collectors are much stiffer than the electrodes. We also assume that the thicknesses in the two layers remains comparable. Note that, in general, if the stiff layer is also much thinner than the soft layer, one has to be careful: a stiff but thin layer may appear to be more compliant than a soft but thick layer. In the case of batteries, the ratio in thickness between electrode and CC is less than a factor of ten, while the difference in stiffness about two orders of magnitude. Therefore, we look at the limit in which the CC are much stiffer than the electrodes. Letting the small parameter $\xi = \frac{\tilde{E}_A}{\tilde{E}_-}\ll 1$, we write $E_- = \bar{E}_-/ \xi$ and $E_+=\bar{E}_+/\xi$ where $\bar{E}_-$, $\bar{E}_+$, $E_A$ and $E_C$ are all of order one. We expand all dependent fields as a series in small $\xi$, 
\begin{align}
    T^A_{\eta \zeta}&= T^{A (0)}_{\eta \zeta} +T^{A (1)}_{\eta \zeta}\xi+T^{A (2)}_{\eta \zeta} \xi^2+...\\
    T^C_{\eta \zeta}&=T^{C (0)}_{\eta \zeta} +T^{C (1)}_{\eta \zeta} \xi+T^{C (2)}_{\eta \zeta} \xi^2+...\\
    T^{\pm}_{\eta \zeta}&=T^{\pm (-1)}_{\eta \zeta}\xi^{-1}+T^{{\pm} (0)}_{\eta \zeta} +T^{{\pm} (1)}_{\eta \zeta} \xi+..., \quad \eta, \zeta \in \{2,3\}.
\end{align}
where now the superscript $(i)$ refers to the expansion in $\xi$ and not anymore in $\delta$. Note that the lowest order terms in the expansion of $T^{\pm}_{\eta \zeta}$ are of order $\xi^{-1}$ since they are proportional to $E_-$ and $E_+$ which are of order $\xi^{-1}$. We also expand the in-plane displacements (these are the same in the CC and their respective electrode, from \eqref{condition_u}) as
\begin{align} u^C_{\eta}=u^+_{\eta}&= u^{+(0)}_{\eta} + u^{+(1)}_{\eta} \xi +u^{+(2)}_{\eta} \xi^2+ \dots\\
    u^A_{\eta}=u^-_{\eta}&=   u^{-(0)}_{\eta} + u^{-(1)}_{\eta} \xi +u^{-(2)}_{\eta} \xi^2+ \dots,\quad \eta \in \{2,3\},
\end{align}
and the through-cell displacements in the layers and $\Delta u^k_1$ as
\begin{align}
     u^k_{1}&= u^{k(0)}_1 + u^{k(1)}_{1} \xi +u^{k(2)}_{1} \xi^2+ \dots,\\
    \Delta u^k_{1}&=  \Delta u^{k(0)}_1 +  \Delta u^{k(1)}_{1} \xi + \Delta u^{k(2)}_{1} \xi^2+ \dots .
\end{align}

\subsection{Leading-order solution: no strain condition}
We begin by expanding the equilibrium equations \eqref{eq:equilibriumF}--\eqref{bc:side2B} in powers of $\xi$. At leading order we need to solve
\begin{align}
\nabla \cdot \tensor{T}^{\pm (-1)}&=0,
\end{align}
with boundary condition
\begin{align}
    \tensor{T}^{\pm (-1)} \cdot \vec{\hat{x}}_2&=0 \quad \text{at} \,\, y= \pm \half L,\\
\tensor{T}^{\pm (-1)} \cdot \vec{\hat{x}}_3&=0 \quad \text{at} \,\, z= \pm \half W,
\end{align}
together with equation \eqref{equation_u1} for the through-cell stress. Using equation \eqref{eq:tension_leading_order}, our problem for the displacements reduces to solving
\begin{align}
\notag
  \frac{2}{(1-2 \nu_{\pm})} \left((1-\nu_{\pm}) \frac{\partial^2 u_2^{\pm(0)}}{\partial x_2^2} + \nu_{\pm} \left( \frac{\partial^2 u_3^{\pm(0)}}{\partial x_2 \partial x_3} + \frac{1}{t_k}\frac{\partial \Delta u_{1}^{\pm(0)}}{ \partial x_2} \right) \right)+
 \left( \frac{\partial^2 u_2^{\pm(0)}}{\partial x_3^2} + \frac{\partial^2 u_3^{\pm(0)}}{\partial x_2 \partial x_3} \right)& =0,\\
  \frac{2}{(1-2 \nu_{\pm})} \left((1-\nu_{\pm}) \frac{\partial^2 u_3^{\pm(0)}}{\partial x_3^2} + \nu_{\pm} \left( \frac{\partial^2 u_2^{\pm(0)}}{\partial x_2 \partial x_3} +\frac{1}{t_k}\frac{\partial \Delta u_{1}^{\pm(0)}}{ \partial x_3} \right) \right)+
 \left( \frac{\partial^2 u_3^{\pm(0)}}{\partial x_2^2} + \frac{\partial^2 u_2^{\pm(0)}}{\partial x_2 \partial x_3} \right)& =0,\\
\p{u_{1}^{\pm(0)}}{x_1}+\frac{\nu_{\pm}}{1+\nu_{\pm}}\left(\p{u_{2}^{\pm(0)}}{x_{2}}+\p{u_{3}^{\pm(0)}}{x_{3}}\right)&=0,
\end{align}
with boundary conditions
\begin{align}
    \p{u_2^{(0)\pm}}{x_3}+ \p{u_3^{(0)\pm}}{x_2}=0, \quad \p{u_2^{(0)\pm}}{x_2}(1-\nu_{\pm})+\nu_{\pm} \left(\frac{\Delta u_1^{\pm}}{t_k}+\p{u_3^{(0)\pm}}{x_3}\right)=0 \quad \text{at} \,\, y= \pm \half L,\\
    \p{u_2^{(0)\pm}}{x_3}+ \p{u_3^{(0)\pm}}{x_2}=0, \quad \p{u_3^{(0)\pm}}{x_3}(1-\nu_{\pm})+\nu_{\pm} \left(\frac{\Delta u_1^{\pm}}{t_k}+\p{u_2^{(0)\pm}}{x_2}\right)=0=0 \quad \text{at} \,\, z= \pm \half W.
\end{align}
Since in the current collector there is no swelling term $\alpha$ driving the system, the leading-order problem is solved by setting all the in-plane displacements, as well as the  through-cell stresses in the CC, to vanish:
$u_{\eta}^{A(0)}=u_{\eta}^{-(0)}=u_{\eta}^{C(0)}=u_{\eta}^{+(0)}=0$ and $u_1^{\pm(0)}=0$. This means that the only displacement is in the electrodes and only in the through-cell direction.

\subsection{The leading-order stress in the electrode}
Since the stiff current collectors only allow through-cell displacement, the leading-order stresses in the electrodes, given by equations \eqref{stressExpansion2} and \eqref{stressExpansion3}, read:
\begin{align}
     \sigma^{k(0)}_{11}&=\frac{E_k}{(1-2\nu_k)}\left[ \frac{(1-\nu_k)}{(1+ \nu_k)} \left(\p{u_1^{k(0)}}{x_1}\right)  - \alpha_k \right] \delta_{\eta \zeta}, \\
     \sigma^{k(0)}_{\eta \zeta }&=\frac{E_k}{(1-2\nu_k)}\left[ \frac{\nu_k}{(1+ \nu_k)} \left(\p{u_1^{k(0)}}{x_1}\right)  - \alpha_k \right] \delta_{\eta \zeta}.
\end{align}
where $k \in\{A,C\}$.
Similarly, the leading-order through-cell equilibrium equations coming from \eqref{equation_u1} read
\begin{align}
    \p{u^{A(0)}_1}{x_1}&= \frac{1+\nu_A}{1-\nu_A}\alpha_A+\frac{(1-2\nu_A)(1+\nu_A)}{(1-\nu_A)E_A}c(x_2,x_3),\\
    \p{u^{C(0)}_1}{x_1}&= \frac{1+\nu_C}{1-\nu_C}\alpha_C+\frac{(1-2\nu_C)(1+\nu_C)}{(1-\nu_C)E_C}c(x_2,x_3).
\end{align}
Integrating over the thickness, we obtain the total through-cell displacement in each layer:
\begin{align}
\label{u1_sol}
    \Delta u^{A(0)}_1&= \frac{1+\nu_A}{1-\nu_A}t_A \langle \alpha_A \rangle +\frac{(1-2\nu_A)(1+\nu_A)}{(1-\nu_A)E_A} t_A c(x_2,x_3),\\
    \Delta u^{C(0)}_1&= \frac{1+\nu_C}{1-\nu_C} t_C \langle \alpha_C \rangle +\frac{(1-2\nu_C)(1+\nu_C)}{(1-\nu_C)E_C} t_C c(x_2,x_3),
\end{align}
where $\langle \alpha_k \rangle$ is defined in \eqref{thicknessaverage}.
We can also find the total change in thickness of the electrodes between the two layers, given by $\Delta u= \Delta u^{A(0)}_1+\Delta u^{C(0)}_1$ (note that $\Delta u$ is constant since the two current collectors remain parallel and flat). This allows us to write:
\begin{align}
\label{eq:c}
 \sigma^{(0)}_{11}(x_2,x_3) = c(x_2,x_3) = \frac{\Delta u - \langle \alpha_A \rangle \left(\frac{1+\nu_A}{1-\nu_A}\right)t_A-\langle \alpha_C \rangle \left(\frac{1+\nu_C}{1-\nu_C}\right)t_C}{\frac{(1+\nu_A)(1-2 \nu_A)t_A}{(1-\nu_A)E_A}+\frac{(1+\nu_C)(1-2 \nu_C)t_C}{(1-\nu_C)E_C}}.
\end{align}
Finally, if we are applying an external load, we can find $\Delta u$ using equation \eqref{bc:global}, where
\begin{align}
\label{bc:global2}
  \int_0^{L} \int_0^{W} c(x_2,x_3)\; dx_2 dx_3 &=N=-p L W,
\end{align}
leading to:
\begin{align}
\Delta u =  t_A \left(\frac{1+\nu_A}{1-\nu_A}\right) \overline{\langle \alpha_A \rangle}+t_C \left(\frac{1+\nu_C}{1-\nu_C}\right) \overline{\langle \alpha_C \rangle}-\left({\frac{(1+\nu_A)(1-2 \nu_A)t_A}{(1-\nu_A)E_A}+\frac{(1+\nu_C)(1-2 \nu_C)t_C}{(1-\nu_C)E_C}}\right) p.
\end{align}
The in-plane stresses in the electrodes are given by
\begin{align}
\sigma^{k(0)}_{\eta \zeta}=\left[\frac{\nu_k}{(1-\nu_k)} c(x_2,x_3)-\frac{E_k}{1-\nu_k}  \alpha_k  \right]\delta_{\eta \zeta}, \quad \eta, \zeta \in \{2,3\}.
\label{eqn:stresses_electrode}
\end{align}
Equations \eqref{eq:c} and \eqref{eqn:stresses_electrode} are the main results in this paper, and entirely describe the stress field in the battery's electrodes.

\subsection{Tension in the current collectors}
Although we have solved the problem in the soft layer, it is useful to also determine the resulting tension in the stiff current collectors. This tension can be evaluated looking at the next order expansion in $\xi$ of the equilibrium equations,
\begin{align}
    \label{eq:TA}
    \nabla \cdot ( \half \tensor{T}^{-(0)}+\tensor{T}^{A (0)})&=0,\\
    \label{eq:TC}
    \nabla \cdot ( \half \tensor{T}^{+(0)}+\tensor{T}^{C (0)})&=0,
\end{align}
with boundary conditions \eqref{bc:side1B} and \eqref{bc:side2B}. We re-write these boundary conditions more generally over a given (not necessarily rectangular) boundary domain. Letting the domain be $\Omega$ and its boundary $\delta \Omega$ with normal $\vec{n}$, the boundary conditions are
\begin{align}
\label{bc:TA}
    (\half \tensor{T}^{-(0)}+\tensor{T}^{A (0)}) \cdot \vec{n}&=0,\\
    \label{bc:TC}
     (\half \tensor{T}^{+(0)}+\tensor{T}^{C (0)}) \cdot \vec{n}&=0 \quad \vec{x} \in \delta \Omega.
\end{align}

Importantly, since we have solved the leading-order problem in the electrode layers, we know that the tension in the $A$ and $C$ electrode  is
\begin{equation}
T^{s (0)}_{\eta \zeta} \equiv - \half  f_s(x_2,x_3)\delta_{\eta
  \zeta}=- \left[\frac{E_s t_s}{1-\nu_s} \langle \alpha_s
  \rangle-\frac{\nu_s}{1-\nu_s}c(x_2,x_3) t_s \right] \delta_{\eta
  \zeta}, \qquad s \in \{A,C\}.
\end{equation}
Therefore, equilibrium reduces to
\begin{align}
\label{eq:TAB}
\nabla \cdot \tensor{T}^{-(0)}&=\nabla f_A(x_2,x_3),\\
\label{eq:TCB}
\nabla \cdot \tensor{T}^{+(0)}&=\nabla f_C(x_2,x_3),
\end{align}
where $f_A$ and $f_C$ are known functions coming from the lower order solution.

Before we proceed to solve the above problem, we recall that the divergence of the tension in the CC is just the leading-order shear-stress at the interface between CC and electrode, as obtained in equation  \eqref{eq:equilibriumO1}. Therefore, we deduce that the leading-order shear stresses at the interface between CC and electrode (i.e. at $x=-t_A$ and $x=t_C$ respectively) are given by:
\begin{align}
\label{eq:shears1}
    \sigma^A_{1 \zeta}&=  \half \delta \,(\partial_{\eta} f_A), \quad  \text{at} \quad x_1=-t_A,\\
    \label{eq:shears2}
    \sigma^C_{1 \zeta}&=  \half \delta \,(\partial_{\eta} f_C ) \quad  \text{at} \quad x_1=t_C.
\end{align}
Note that we have expressed explicitly the $\delta$ dependence here since shears are $O(\delta)$ in our thin-limit expansion.

\subsection{The tension in the current collector using an Airy stress function}
Finally, we proceed to solve the problem for the tension in the current collectors. We use the Airy stress function $\phi$ to simplify the problem letting:
\begin{align}
\label{eq:sigma_def}
\hat{\Sigma}^{\pm}_{22}=  \frac{\partial^2 \phi^{\pm}}{\partial x_3^2},\quad
\hat{\Sigma}^{\pm}_{33}= \frac{\partial^2 \phi^{\pm}}{\partial x_2^2},\quad  
\hat{\Sigma}^{\pm}_{23}=- \frac{\partial^2 \phi^{\pm}}{\partial x_2 \partial x_3}.
\end{align}
By construction, we have that
\begin{align}
\label{Tresc}
 \hat{T}^{{+}(0)}_{22}&= \frac{\partial^2 \phi^{+}}{\partial x_3^2}+f_C, &\quad \hat{T}^{{-}(0)}_{22}&= \frac{\partial^2 \phi^{-}}{\partial x_3^2}+f_A,\\
 \hat{T}^{{+}(0)}_{33}&= \frac{\partial^2 \phi^{+}}{\partial x_2^2}+f_C, &\quad  \hat{T}^{{-}(0)}_{33}&= \frac{\partial^2 \phi^{-}}{\partial x_2^2}+f_A,\\ 
 \hat{T}^{{+}(0)}_{23} &=-\frac{\partial^2 \phi^{+}}{\partial x_2 \partial x_3},& \quad \hat{T}^{{-}(0)}_{23} &=-\frac{\partial^2 \phi^{-}}{\partial x_2 \partial x_3}.
\end{align}
The tensions in the CC must satisfy a stress compatibility condition, arising from the compatibility of strain, given by:
\begin{align}
\label{eqn:stressCompatibility}
\frac{\partial^2 \hat{T}^{+(0)}_{22}}{\partial x_3^2} + \frac{\partial^2 \hat{T}^{+(0)}_{33}}{\partial x_2^2} - \nu_C \left(\frac{\partial^2 \hat{T}^{+(0)}_{22}}{\partial x_2^2} + \frac{\partial^2 \hat{T}^{+(0)}_{33}}{\partial x_3^2}\right) - 2 (1+\nu_C) \frac{\partial^2 \hat{T}^{+(0)}_{23}}{\partial x_2 \partial x_3} &= 0,\\
\frac{\partial^2 \hat{T}^{-(0)}_{22}}{\partial x_3^2} + \frac{\partial^2 \hat{T}^{-(0)}_{33}}{\partial x_2^2} - \nu_A \left(\frac{\partial^2 \hat{T}^{-(0)}_{22}}{\partial x_2^2} + \frac{\partial^2 \hat{T}^{-(0)}_{33}}{\partial x_3^2}\right) - 2 (1+\nu_A) \frac{\partial^2 \hat{T}^{-(0)}_{23}}{\partial x_2 \partial x_3} &= 0
\end{align}
leading to the bi-harmonic problems:
\begin{align}
    \nabla^4 \phi^+ + \half (1-\nu_C)\nabla^2 f_C=0, \quad \quad
    \nabla^4 \phi^- + \half (1-\nu_A)\nabla^2 f_A=0.
    \label{eqn:biham_final}
\end{align}
Note that, since we have already solved for the functions $f_A$ and
$f_C$ at leading order, the problems in the electrodes and the CCs are
completely decoupled (unlike when the CCs where not stiff) and can be solved independently. 
Solutions to equations \eqref{eqn:biham_final} need to satisfy specific boundary conditions.  

The boundary condition on $\phi^{\pm}$ come from  equations \eqref{bc:TA} and \eqref{bc:TC} and, in general, require $\hat{\tensor{\Sigma}}^{\pm} \cdot \vec{n}=0$ at the boundary, where $\vec{n}$ is normal to the boundary. For a general $n=(n_2,n_3)$, this condition reads
\begin{equation}
\hat{\tensor{\Sigma}}^{\pm} \cdot \vec{n} = 
\begin{pmatrix}
    \frac{\partial^2 \phi^{\pm}}{\partial x_3^2} && -\frac{\partial^2 \phi^{\pm}}{\partial x_2 \partial x_3} \\
    -\frac{\partial^2 \phi^{\pm}}{\partial x_2 \partial x_3} && \frac{\partial^2 \phi^{\pm}}{\partial x_2^2}
\end{pmatrix}
\begin{pmatrix}
    n_2 \\
    n_3
\end{pmatrix}
=
\begin{pmatrix}
    -\frac{\partial^2 \phi^{\pm}}{\partial x_2 \partial x_3} n_3 + \frac{\partial^2 \phi^{\pm}}{\partial x_3^2} n_2\\
    \frac{\partial^2 \phi^{\pm}}{\partial x_2^2} n_3 - \frac{\partial^2 \phi^{\pm}}{\partial x_2 \partial x_3} n_2
\end{pmatrix}
=
\begin{pmatrix}
    0 \\
    0
\end{pmatrix}.
\end{equation}

This can be simplified significantly if we first recognise the tangent to the boundary is $\vec{t}=(n_3,-n_2)$, then we can equivalently write the boundary condition as
\begin{equation}
\nabla \nabla \phi^{\pm} \cdot \vec{t} =
\begin{pmatrix}
    \frac{\partial^2 \phi^{\pm}}{\partial x_2^2} && \frac{\partial^2 \phi^{\pm}}{\partial x_2 \partial x_3} \\
    \frac{\partial^2 \phi^{\pm}}{\partial x_2 \partial x_3} && \frac{\partial^2 \phi^{\pm}}{\partial x_3^2}
\end{pmatrix}
\begin{pmatrix}
    n_3 \\
    -n_2 \\
\end{pmatrix}
=
\begin{pmatrix}
    \frac{\partial^2 \phi^{\pm}}{\partial x_2^2} n_3 - \frac{\partial^2 \phi^{\pm}}{\partial x_2 \partial x_3} n_2\\
    \frac{\partial^2 \phi^{\pm}}{\partial x_2 \partial x_3} n_3 - \frac{\partial^2 \phi^{\pm}}{\partial x_3^2} n_2
\end{pmatrix}
=
\begin{pmatrix}
    0 \\
    0
\end{pmatrix}
\end{equation}
where the only difference is the sign of the first entry in the vector. We then integrate $\nabla \nabla \phi^{\pm} \cdot \vec{t}$ along the boundary from an arbitrary point $\vec{a}$ to an arbitrary  point $\vec{b}$ to find 
\begin{equation}
\int_a^b \nabla \nabla \phi^{\pm} \cdot \vec{t}\; ds= \nabla \phi^{\pm}(\vec{b})-\nabla \phi^{\pm}(\vec{a})=0.
\end{equation}
Since $\vec{a}$ and $\vec{b}$ are arbitrary, $\nabla \phi^{\pm}$ must be a constant vector over the boundary. Now, suppose the constant is different from zero, say equal to the vector $\vec{v}=(v_2,v_3)$. We can always subtract from $\phi^{\pm}$ the field $\phi^{\pm}_v=v_2 y+v_3 z$ and get $\nabla \phi^{\pm}= 0$ without affecting the stresses. As a result, we can always set the gradient to vanish on the boundary.
Letting $\nabla \phi^{\pm}=0$ on the boundary means that $\phi^{\pm}$ must be constant too. We thus have a second gauge freedom that allows us to set $\phi^{\pm}=0$. Finally, in summary, the boundary conditions for our problem \eqref{eqn:biham_final} are
\begin{equation}
\label{bc:bham}
    \nabla \phi^{\pm}(\vec{x})= \vec{0}\quad\ \text{and}\quad\ \phi^\pm(\vec{x})=0\quad\ \text{for}\,\,\,\vec{x}\in \delta \Omega.
\end{equation}

Taking the limit in which the CC becomes stiff we have therefore found explicit expressions for the stress in the electrode and the shear stress at the boundary between current collector and electrode. Furthermore, using an Airy stress function, we have derived a bi-harmonic problem that describes the tension in the CC, in which the lithiation expansion $\alpha$ acts as a source term. 

\subsection{Homogeneous expansion:}

Having solved the general problem for thin layers and stiff current collectors, we can explicitly write the solution for the simple case of homogeneous expansion, where $\alpha_C$ and $\alpha_A$ are constants independent of space. If we consider a system subject to an external loading, the problem is effectively one dimensional. The through-cell stress is given by
\begin{align}
\label{eq:c_hom}
 \sigma^{(0)}_{11}=N/LW=-p.
\end{align}
Conversely, if we clamp the structure so that we control $\Delta u$, our $\sigma^{(0)}_{11}$ is given by:
\begin{equation}
\label{eq:c_hom_disp}
\sigma^{(0)}_{11}=\frac{\Delta u - \alpha_A \left(\frac{1+\nu_A}{1-\nu_A}\right)t_A- \alpha_C \left(\frac{1+\nu_C}{1-\nu_C}\right)t_C}{\frac{(1+\nu_A)(1-2 \nu_A)t_A}{(1-\nu_A)E_A}+\frac{(1+\nu_C)(1-2 \nu_C)t_C}{(1-\nu_C)E_C}}.
\end{equation}
In either case, the in-plane stresses in the electrodes are isotropic and are given by 
\begin{align}
\label{eqn:stresses_electrode_hom}
     \sigma^{s(0)}_{\eta \zeta}&=- \frac{1}{{1-\nu_s}}\left(E_s  \alpha_s -\nu_s \sigma_{11}^{(0)} \right)\delta_{\eta \zeta} \quad s \in \{A,C\}.
\end{align}
The tension in the current collectors simplifies dramatically when the expansion is homogeneous and can be expressed in closed form. In such case, the source term in the biharmonic equation \eqref{eqn:biham_final} vanishes, and the biharmonic problem is solved by $\phi\equiv 0$ everywhere, so that the tension is simply given by:
\begin{align}
\label{eq:tension_hom}
    T^{\pm(0)}_{\eta \zeta}=t_s \sigma^{s(0)}_{\eta \zeta}= \frac{2 t_s}{{1-\nu_s}}\left(E_s  \alpha_s -\nu_s \sigma_{11}^{(0)}  \right)\delta_{\eta \zeta} \quad s \in \{A,C\}.
\end{align}
The fact that the solution has rotational symmetry (it is a multiple of identity) reflects that our problem is effectively one dimensional.

We remark that solving the homogeneous expansion problem did not require us to exploit the thin layer limit, since the displacements are only in the $x_1$ direction and have no spatial variation. However, the stiff CC limit is important, as otherwise a more complex solution would arise with non-zero in-plane displacements. 

\section{Comparison with 3D FE simulations}

\begin{figure}[t]
    \centering
    \includegraphics[width=\textwidth]{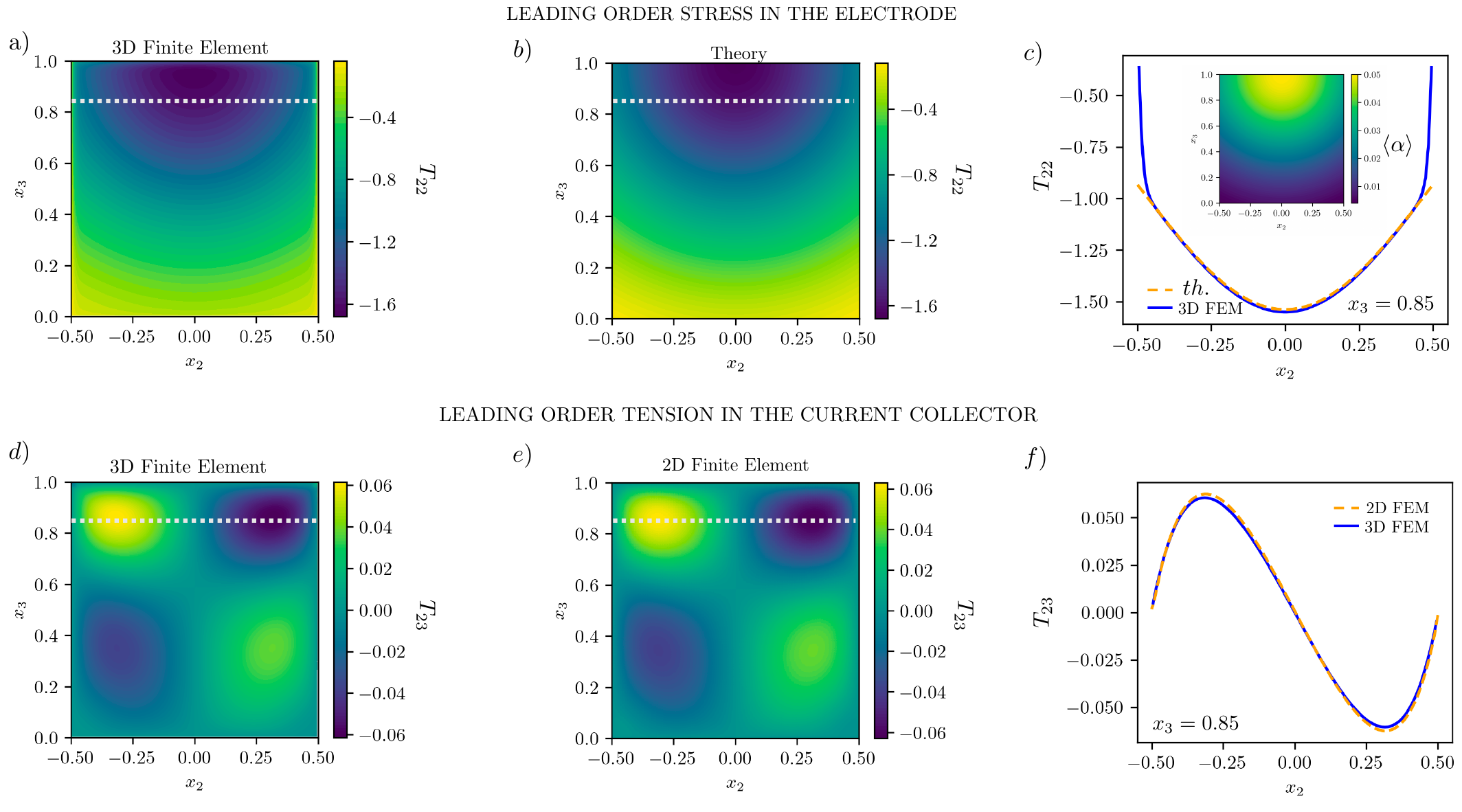}
    \caption{Comparison between theory and full 3D finite element simulations of the tension in an electrode and current collector pair. Our results are presented in dimensionless form. The 3D FE simulations was carried out in FeBio. In the
      top row, panels a), b), we compare the $x_2$ component
      of the tension in the electrode over the whole $x_2$,
      $x_3$ plane while in c) we compare this tension along the dashed
      grey line at $\tilde{x}_3=0.85$. In the lower row, panels d) and
      e), we compare the shear stress distribution in the electrode
      while in f) we compare this shear stress along the dashed grey line at
      $x_3=0.85$. Note that the $x_2$ and $x_3$
      axis are dimensionless.}
    \label{fig:numerics}
\end{figure}

We now compare our theoretical results with full 3D Finite Element
simulations. The FE simulations are carried out in the open source
software FeBio \citep{maas2012febio}, designed for biological systems
and thus able to efficiently simulate expanding (or pre-strained)
linear elastic materials. Unfortunately, simulating the behaviour of
the thin and compliant separator is complicated in the FE
simulations. This is because the separator may undergo large
deformations and shears, leading to convergence problems. To avoid
this complication, we choose to simulate a simpler system in which
both cathode and anode swell by the same amount locally, and with the
same material properties, i.e. the separator acts as a symmetry
plane. Note that this is not a realistic battery simulation, but has the intention of verifying the results from the previous sections, before we simulate more carefully the stresses in a battery in the next subsection.

\begin{table}[h]
\begin{center}
\renewcommand{\arraystretch}{1.5} 
\begin{multicols}{2}
    \begin{tabular}{|c c|} 
    \hline
    Parameter & Value  \\ [0.2ex] 
    \hline\hline
      $\delta$ & $10^{-2}$ \\  [0.1ex]
      \hline
       $\xi$ & $10^{-3}$ \\  [0.1ex]
     \hline
     $t_A$ & $1 $ \\ [0.1ex]
      \hline
     $t_-$ & $1 $ \\  [0.1ex]
      \hline
       $L$ & $1 $ \\  [0.1ex]
       \hline
    \end{tabular}

    \columnbreak
    
    \begin{tabular}{|c c|} 
    \hline
    Parameter & Value  \\ [0.2ex] 
    \hline\hline
         $W$ & $1 $ \\ [0.1ex]
      \hline
      $E_A=\overline{E}_-$ & $1$ \\  [0.1ex]
     \hline
     $E_-=\overline{E}_-/\xi$ & $ 10^3$ \\  [0.1ex]
     \hline
     $\nu_A$ & $0.2$  \\ [0.1ex]
     \hline
     $\nu_-$ & $0.2$  \\  [0.1ex]
     \hline
    \end{tabular}
\end{multicols}
\end{center}
\caption{Dimensionless parameter values used in FEBio for the finite element simulations of an anode and ($+$) CC.}
\label{table1}
\end{table}

The proposed simplification allows us to run our 3D finite element simulations on just one CC with its electrodes (on either side). Due to the symmetry about the CCs mid-plane, we can simplify this further  and simulate only one electrode attached to half of its current collector, requiring both surfaces normal to to the through-cell direction to remain planar. The simulations are run assuming the electrode (anode) and CC ($-$) have the same thickness $t_A=t_-=1$, and the domain is a square with $L^*=100 t^*$ meaning $\delta=t^*/L^*=0.01$ and $L=W=1$. The ratio in stiffness (Young's modulus) between the electrode (anode) and the current collector ($-$) is set to $\xi=0.001$ and the Poisson ratio to $\nu_A= \nu_- = 0.2$. A list of all the dimensionless parameters is shown in table \ref{table1}.
The electrode domain is meshed with a $10 \times 100 \times 100$ grid for the electrode and a $5 \times 100 \times 100$ for the current collector. This has proved to be sufficient for convergence of our results. As boundary condition, we imposed the $x_1$ displacement to vanish on the two free surfaces normal to $\vec{\hat{x}}_1$ (thus implying a global clamping condition $\Delta u=0$). We choose an arbitrary expansion function aimed at generating a sufficiently complex stress field to verify our analytic results, we let
\begin{equation}
    \alpha = \exp\left(-\frac{2y^2+2(z-\half L)^2}{L^2} \right)\left(\frac{13}{12} - \left(x - \frac{1}{2} \right)^2 \right). 
\end{equation}
Note that the quantity in the second bracket integrated over the thickness is one, so that the maximum amplitude averaged through the thickness remains one. The profile of swelling (averaged over the $x_1$ axis) is shown in the inset of figure \ref{fig:numerics}c).

In figure \ref{fig:numerics} a), b) and c) we compare the FE results in the electrode with theoretical predictions from \eqref{eqn:stresses_electrode}. To better present the results, we have integrated \eqref{eqn:stresses_electrode} through the thickness and obtained the tension $T_{22}=T_{33}$. 
Figure \ref{fig:numerics} d), e) and f) instead shows results for shear component of tension in the current collector. We solved the bi-harmonic problem in \eqref{eqn:biham_final} using the finite element solver Dolfinx in Python and, once more, compared the 2D results with the $x_1$- averaged results from the full 3D finite elements. In both cases, our theoretical predictions are in excellent agreement with the 3D FE model, with small discrepancy only near the boundary.

\subsection{Boundary layer effects}

In figure \ref{fig:numerics}c) it is apparent that our asymptotic solution predicts well both stresses and tensions in a battery as long as we are not too close to the boundary. This is because, as discussed in section \ref{sec:3Dproblem_setup}, there is a boundary layer where the materials (and stresses) partially relax. On dimensional grounds, we know that the boundary region scales like the thickness of the layer and is thus small compared to the size of the battery, which is why our bulk solution does such a good job in a system in which $\delta \sim 1/100$. To resolve this boundary layer, further local analysis --- which is beyond the scope of this work --- is required.

\subsection{Stress distribution for a discharging battery}

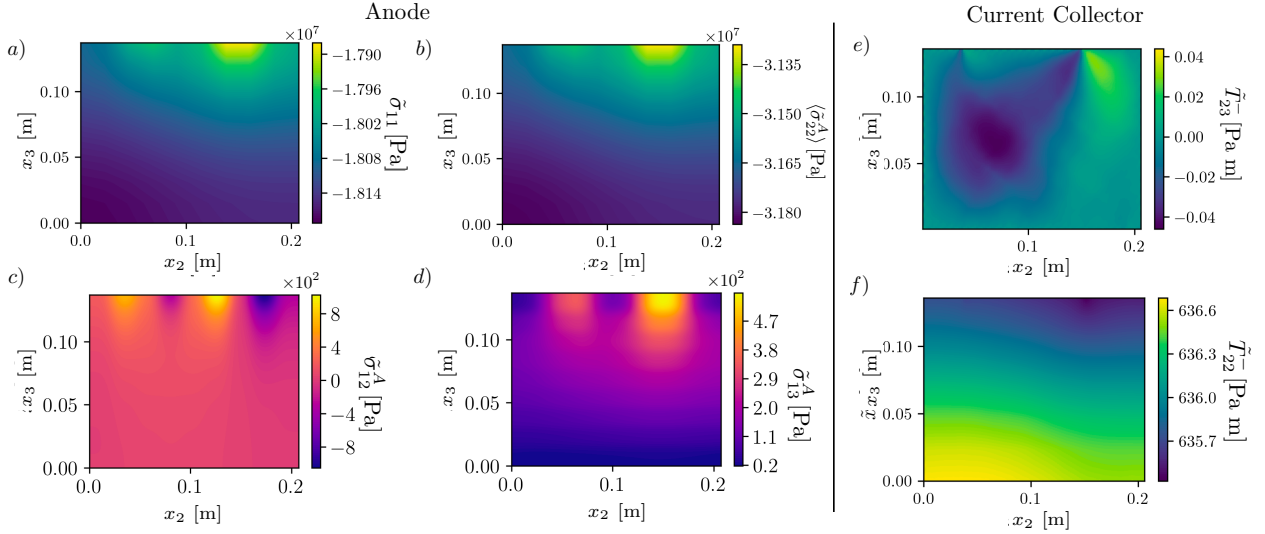
\begin{figure}[t]
  \centering
  \begin{overpic}[width=\textwidth]{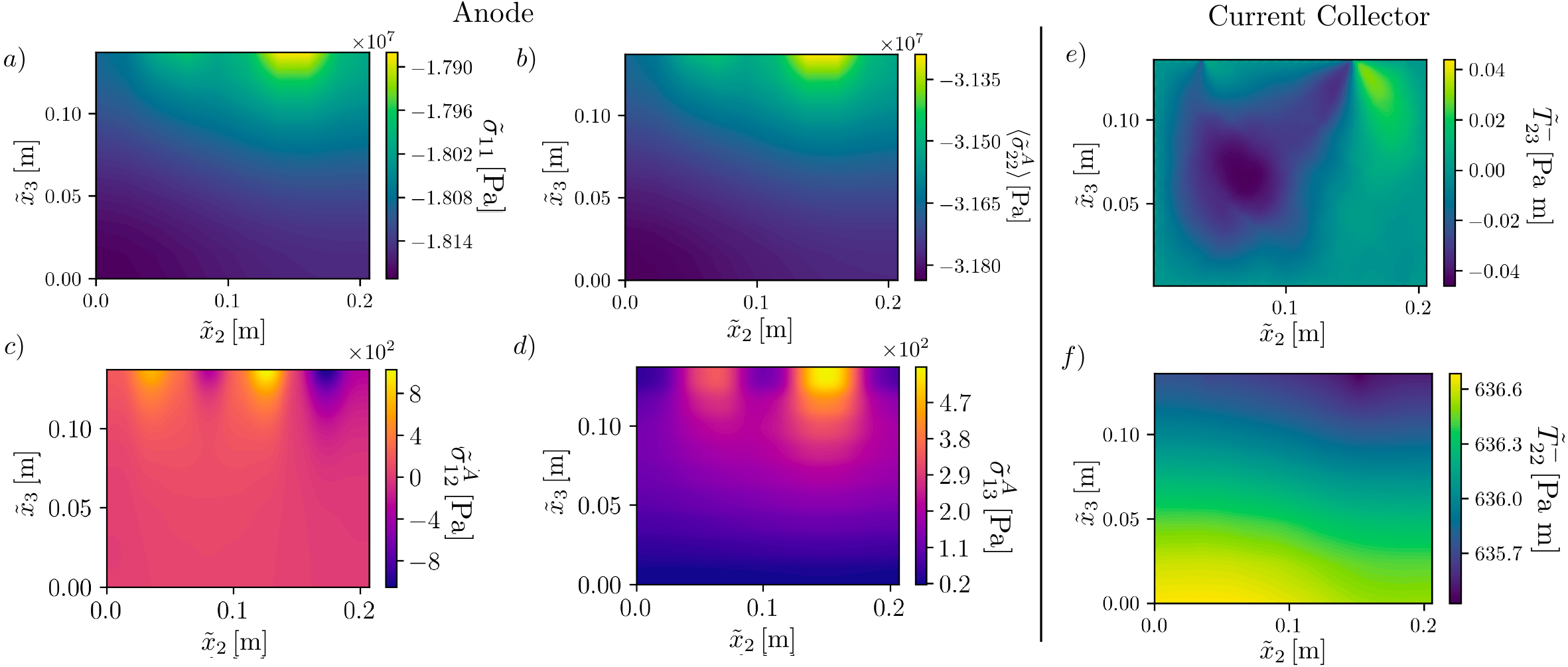}
    \put(0.8,29){\hl{\rotatebox{90}{\scriptsize $x_3$ [m]}}} 
    \put(0.8,10){\hl{\rotatebox{90}{\scriptsize $x_3$ [m]}}} 
    \put(34.5,10){\hl{\rotatebox{90}{\scriptsize $x_3$ [m]}}} 
    \put(34.5,29){\hl{\rotatebox{90}{\scriptsize $x_3$ [m]}}} 
    \put(68.5,10){\hl{\rotatebox{90}{\scriptsize $x_3$ [m]}}} 
    \put(68.5,29){\hl{\rotatebox{90}{\scriptsize $x_3$ [m]}}} 
    \put(12.5,21.2){\hl{\scriptsize $x_2$ [m]}} 
    \put(12.5,1){\hl{\scriptsize $x_2$ [m]}} 
    \put(46.5,21.2){\hl{\scriptsize $x_2$ [m]}} 
    \put(46.5,1.2){\hl{\scriptsize $x_2$ [m]}} 
    \put(80.5,20.8){\hl{\scriptsize $x_2$ [m]}} 
    \put(80.5,0.5){\hl{\scriptsize $x_2$ [m]}} 
\end{overpic}
    \caption{ Simulations of a battery discharging at 2C using PyBaMM. The anode and cathode have the same elastic modulus $10^9$ Pa and Poisson ratio of $\nu = 0.2$. Simulations are shown half way through the discharging process (i.e. after 15 minutes) for the anode and $(-)$ CC as this is where the largest effects are observed, with similar but milder ones in the other pair. Panel a) shows the through-cell stress, b) the in-plane stress state averaged over the thickness, c) and d)  show the shear stress at the interface between anode and current collector obtained from equations \eqref{eq:shears1} and \eqref{eq:shears2}. Panels e) and f) show the $23$ and $22$ component of tension. The gradients are a result of the uneven current flow which is largest near the tabs. In our simulations, the tabs are at the top of the system ($x_3=0.137)$ and correspond with the locations of the largest through-cell stress in b).}
    \label{fig:batterysim}
\end{figure}

We now show how the results obtained in section \ref{sec:stifflimit} of this paper can predict the stress state in a discharging lithium-ion battery. Note that we shall predict the behaviour of the battery neglecting how the stress couples to the electrochemistry. That is, we first solve the electrochemical problem and then compute the stresses that would occur. We intend to examine the coupled problem, where the stress affects the electrochemical behaviour, in a later paper. 

We simulate the electrochemical state of a pouch cell using the python
battery simulation package PyBaMM \citep{sulzer2021python}.  
We make two further simplifying assumptions: first, we make the realistic choice that the battery is clamped so that its thickness cannot change and $\Delta u = 0$. Furthermore, we assume that both electrodes have the same stiffness $\tilde{E}_A=\tilde{E}_C=1$ GPa and Poisson ratio of $\nu=0.2$, consistent with previously reported quantities \citep{uccel2022experimental}.

We run a simulation using potential pair current collectors proposed
by \cite{timms2021asymptotic}. This model has a lumped through-cell
direction, but accounts for in-plane variations in the $x_2$-$x_3$
directions, allowing us to  capture inhomogeneities across the electrodes. We use the parameters from \citep{marquis2019asymptotic} with both electrodes of equal thickness $10^{-4}$ m. However, to exaggerate the inhomogeneities produced during the discharging process, we set the current collectors to have thickness $10^{-5}$ m. We then discharge an initially fully charged battery at a relatively fast rate of $2C$.  

The simulation provides us with the electrical state of the battery at any point in the $(x_2,x_3)$ plane and time. For instance, it tells us the stoichiometric coefficient of lithium in the anode and cathode. However, unfortunately, PyBaMM cannot tell us the local expansion in the battery which we would need to determine the stress. Therefore, we need to establish a relationship between the concentration of lithium and the local expansion. To do so, we rely on experimental evidence showing that the intercalation of lithium leads to an overall volumetric expansion of $13.2\%$ in a graphite anode \citep{schweidler2018volume}. The expansion in the cathode is significantly smaller and we shall assume, for the current presentation, that it is vanishingly small. Then, since the volume strain in the anode is $dV/V=3 \alpha_A$ (valid for small strain), we deduce that the maximum swelling is $(\alpha_A)_{max}=4.4 \%$. Assuming a linear relationship between the stoichiometric coefficient $X_n$ --- given by the ratio of lithium ion concentration divided by its maximum possible value --- and the expansion strain, we can set $\alpha_A=4.4\% X_n$. 

The stochiometric coefficient $X_n$ is one of the outputs of the PyBaMM simulation, and thus allows us to turn the electrochemical state of the battery into knowledge of the local expansion. We use this together with equations \eqref{eqn:stresses_electrode} and \eqref{eq:c} to predict the stresses in the battery half way through the discharging process (i.e after 15 minutes). For brevity, we only show the results for the anode layer, where the stresses are largest. The stress state in the cathode layer, generated by the expanding anode, is qualitatively similar, although the magnitude of the in-plane stresses is significantly smaller (about one order of magnitude).

The through-cell and in-plane stress (averaged over the thickness) in the anode are shown in figure \ref{fig:batterysim} a) and b).  We observe a significant compressive stresses of the order of $10^7$ Pa with a spatial inhomogeneity of about $5\%$. This spatial inhomogeneity is a result of the inhomogeneous state of charge of the battery caused by the position of the tabs at the top of the structure (i.e. at $x_3=0.137$). Indeed, near the tabs, the current is slightly larger and causes slightly faster discharge and reduced expansion and swelling as a consequence. The inhomogeneity in expansion causes shear stresses, shown in figure \ref{fig:batterysim} c) and d). The shears are proportional to the thickness of the electrodes and are therefore smaller, of the order of $10^2$-$10^3$ Pa. Finally, in panels e) and f) we show the tension in the negative current collector obtained solving the biharmonic equation \eqref{eqn:biham_final} with boundary conditions \eqref{bc:bham}.

\section{Summary and Conclusion}

During charging and discharging of a lithium-ion pouch cell, the ion-induced swelling of the electrodes, together with the inhomogeneity in mechanical properties of the layers that compose the cell, causes the build up of residual stresses and a macro-scale change in thickness of the battery. In this paper, we have exploited the thin aspect ratio of the battery layers and the large difference in stiffness between electrodes and current collectors to  derive explicit expressions for the through-cell and in-plane stress in the electrodes, given by equations \eqref{eq:c} and \eqref{eqn:stresses_electrode} respectively. Furthermore, we have also derived expressions for the shear stress that is generated at the boundary between current collector and electrode, given in equation \eqref{eq:shears1} and \eqref{eq:shears2}. By defining the tension in the current collectors using a Airy stress function, equations \eqref{eq:sigma_def}, we have derived the biharmonic problem  \eqref{eqn:biham_final} with boundary conditions \eqref{bc:bham}. The  solution to this problem describes the tension in the current collectors. Note that, although all the expressions for the stress have been non-dimensionalised, to obtain their dimensional counterpart one simply substitutes the dimensional form of all the quantities appearing in the equation and sets the scale of swelling $\alpha^*=1$. Our findings are in excellent agreement with full 3D FE simulations. 

In the particular case of a homogeneous expansion, where the swelling is independent of the $x_2$ and $x_3$ coordinates, both the electrode stresses and the tension in the current collectors can be expressed in closed form expressions. The through cell stress is given by equation \eqref{eq:c_hom} in the case of a constant pressure applied on a cell and \eqref{eq:c_hom_disp} in the case of displacement clamping, while the in-plane stress is given in equation \eqref{eqn:stresses_electrode_hom}.  The tension simplifies dramatically, and is given in equation \eqref{eq:tension_hom}. 

Our results for the stress can be used to predict the stress distribution in a charging/discharging battery. We have used PyBaMM, the open-source battery simulation package on Python, to simulate the electro-chemistry of a fast discharging battery. The fast discharge leads to spatial inhomogeneities in the stoichiometry of the battery (i.e. different parts of the battery are charged by different amounts), leading to inhomogeneous expansion in the electrodes and a complex stress field. By converting the local stoichiometric data to an expansion coefficient, and using the expressions for the the stress derived in this work, we are able to simulate the stress-field in the electrodes and current collectors of a discharging battery. This stress-field impacts the micro-scale stress surrounding the active particles which in turn alters the Butler-Volmer equation via the dependence of chemical potential on stress, coupling the macro-scale mechanics with the electro-chemistry.

Finally, we comment on our assumption that a stiff current collector leads to small strains. Given our solution, this can can be checked a posteriori. For simplicity, we consider the homogeneous expansion of the electrode. From the equilibrium the boundary conditions, \eqref{bc:side1} and \eqref{bc:side2}, we know that the average stress in the electrode integrated over the thickness is counterbalanced by the tension in the current collector. From figure \ref{fig:batterysim} b) we observe that an electrode of stiffness $1$ GPa induces a stress in of order $10^7$ Pa. For a current collector of stiffness $100$ GPa and about 1/5 of the electrode  thickness such a stress results in a strain smaller than $10^{-3}$, thus confirming that CCs do not undergo significant deformation.

\section*{Aknowledgments}
This work was generously supported by the EPSRC Faraday Institution
Multi-Scale Modelling project (EP/S003053/1, grant number 
FIRG059)

\appendix

\bibliographystyle{plainnat} 
\bibliography{bib.bib}

\end{document}